\documentclass[twoside,11pt]{article}

\usepackage{uw}

\usepackage[inline]{enumitem}

\ufheading{Semester 1}{2022-2023}{Bib.Review}{Dec.22}{University of Waterloo}{U.Telemaco, P.Alencar, D.Cowan, T.Oliveira}

\ShortHeadings{}{}
\firstpageno{1}

\begin{document}

\title{Agile Assessment Methods: Current State of the Art}

\author{\name Ulisses Telemaco \email utelemac@uwaterloo.ca \\
       \addr David R. Cheriton School of Computer Science \\
       University of Waterloo\\
       Waterloo, Ontario, Canada
       \AND
       \name Paulo Alencar \email palencar@uwaterloo.ca \\
       \addr David R. Cheriton School of Computer Science \\
       University of Waterloo\\
       Waterloo, Ontario, Canada
       \AND
       \name Donald Cowan \email dcowan@uwaterloo.ca \\
       \addr David R. Cheriton School of Computer Science \\
       University of Waterloo\\
       Waterloo, Ontario, Canada
       \AND
       \name Toacy Oliveira \email toacy@cos.ufrj.br \\
       \addr System Engineering and Computing Program \\
       Federal University of Rio de Janeiro\\
       Rio de Janeiro, Brazil}

\maketitle

\begin{abstract}
Agility Assessment (AA) comprises tools, assessment techniques, and frameworks that focus on
indicating how a company or a team is applying agile techniques and eventually pointing out problems in adopting agile practices at a project-level, organization-level or
individual-level. 
There are many approaches for AA such as agility assessment models, agility
checklists, agility surveys, and agility assessment tools. 
This report presents the state of the art approaches that support agility assessment.
\end{abstract}

\begin{keywords}
 Agile Development, Agility Assessment
\end{keywords}

\section{Introduction}

Agile methods use iterative techniques to deliver solutions incrementally by relying on close
collaboration and frequent reassessment. These methods are based on continuous adaptive
planning, development, testing, integration and deployment. They allow organizations to support
a flexible response to change between self-organizing and cross functional teams. Agile methods
capture the specific practices, tactics and tools used to develop software systems that follow the
fundamental principles of Agile development.
The adoption of so-called agile practices may not be straightforward
\citep{fraser2006agile, gandomani2013obstacles, gandomani2013important, qumer2008-agile-adoption, sahota2012-agile-adoption}.
The 15th
Annual State of Agile Survey™ \citep{VersionOne2020} revealed that, although 97\% of companies surveyed claimed
they are using agile practices, only 4
practices. In this scenario, it is important for organizations to identify their gaps in agile
practices, otherwise, the organization may not receive the benefits of adopting them \citep{ambler2012disciplined}.
Agility Assessment (AA) comprises tools, assessment techniques, and frameworks that focus on
indicating problems in adopting agile practices at a project-level, organization-level or
individual-level. There are many approaches for AA such as agility assessment models, agility
checklists, agility surveys, and agility assessment tools.
This report describes the state of the art approaches used for agility assessment. The report
covers


\section{Agility Assessment Approaches}

This section presents a list of agility assessment approaches found in the literature. These
approaches focus on several aspects, including team evaluations, checklists, surveys, metrics,
dashboards, indices, guides, tests, games, maturity levels and matrices, and manifestos.
\\

\noindent
\textbf{1. \textit{42 point test: How Agile are You}}\\
In \cite{waters2008}, the author presented an approach composed of a 42-statement questionnaire that should be used in the following way:
the Project Manager should ask every team member of an agile team (including the product owner, tester, manager, everyone) to review the statements ``honestly''. 
They should score 1 for each statement they believe they are consistent and it could be audited. 
Otherwise they should score 0 for the statement.
The author suggested to calculate the average final score but did not provide any indication of how to analyze this result.
\\
    
\noindent
\textbf{2. \textit{A Better Team}}\\
\cite{hermida2009abetterteam} proposed an online agility assessment approach called \textit{Abetterteam}. The tool has a questionnaire composed of 30 three-option questions. 
The author claimed the tool is able to verify the adoption of the practices proposed by \cite{shore2007art}. 
However, the author did not indicate how the questionnaire is related to the practices proposed by Shore and the rationale behind the assessment result. 
\\

\noindent
\textbf{3. \textit{A Corporate Agile 10-point Checklist}}\\
\cite{yatzeck2012} proposed a two-checklist method to aid the adoption and assessment of agile process in large companies. 
The first checklist is focused on guiding the adoption of Scrum and it is composed of 10 items. 
The second checklist, called ``\textit{You Should Immediately Be Suspicious If}'', describes 8 practices that may indicate misuse of agile practices:
\begin{enumerate*}
    \item \textit{``There is no high-level architecture''},
    \item \textit{``There is no plan''},
    \item \textit{``There is no project dashboard, or you don’t have access''},
    \item \textit{``You aren’t invited to an iteration planning meeting and a showcase for every iteration''},
    \item \textit{``You don’t get any escalations coming out of the planning workshop''},
    \item \textit{``The team performs perfectly in Iteration 1''},
    \item \textit{``You aren’t welcome to join daily standup Scrum meetings as an observer''}, and
    \item \textit{``You can’t get metrics about software quality''}
\end{enumerate*}.
The items in the second checklist are similar to the agile smells proposed in \cite{telemaco2020} since they describe practices that may jeopardize the adoption of agile methods. 
However, these items from the agile smells in some aspects:
\begin{enumerate*}[label=\textit{(\alph*)}]
    \item the practices are described in a generic way and there is no indication of how they could be checked in real scenarios. 
    Therefore, the detection of these practices may be threatened by the bias of the person performing the agility assessment who has to interpret the practice and determine how to check it;
    \item there is no clear relation between the items and the agile practices that motivated them. The author did not explain the origin of the items;
    \item the solution is based on checklists that have to be manually filled by the Project Manager;
    the items are focused on the \textit{Scrum} method.
\end{enumerate*}
\\

\noindent
\textbf{4. \textit{Agile Adoption Interview}} \\
\textit{Agile Adoption Interview} \cite{bonamassa2018AgileAdoptionInterview} is a web-based survey to assist agile team members self-assess their skills in agile development and to provide information about areas of strengths and opportunities for individual improvements.
The survey, which can be used to assess skill in Scrum or Kanban method, is composed of open and closed question organized in 5 sections:
\textit{Overall},
\textit{Team Dynamics},
\textit{Scrum/Kanban Events},
\textit{Scrum/Kanban Intrinsics}, and
\textit{Scrum/Kanban Roles}.
After submitting the responses, the tool sends to the participants an email with the assessment result.
\\

\noindent
\textbf{5. \textit{Agile Alert}} \\
\textit{Agile Alert} \cite{hoffmann2018-AgileAlert} is a web-based assessment tool designed to aid organizations and agile teams to rate their agile capabilities. 
The tool is divided into 2 parts:
part 1, named \textit{``Do you work in an agile framework?''}, is organized in the sections 
\textit{Agile Strategy}, 
\textit{Agile structure}, and
\textit{Agile culture}
while part 2, named \textit{``Are you agile?''}, is organized in the sections 
\textit{Hyperaware},
\textit{Informed Decision-making}, and
\textit{Fast Execution}.
Each section contains statements that describe a specific agile capability and that should be rated using a 5-point Likert scale that ranges from 1 to 5, where 1 means a low capability and 5 a high capability.
The results should be interpreted as follows:
0–30 points: \textit{Agile Alert!};
30–60 points: \textit{Agile Beginner};
60–90 points: \textit{Agile Adopter}; and
90–120 points: \textit{Agile Front-Runner}.
\\
    
\noindent
\textbf{6. \textit{Agile Assessment}} \\
In \cite{nowinski2016-agileassessment}, the author presented a self-assessment agility approach available as a spreadsheet survey with 66 statements/questions grouped in 7 areas: 
\textit{product ownership}, 
\textit{agile process}, 
\textit{team}, 
\textit{quality}, 
\textit{engineering practices},
\textit{fun and learning}, and 
\textit{integration}.
Every team member should assess each statement using a 5 point likert scale. 
The spreadsheet is configured to calculate the average assessment of each statement for the whole team and for each area. 
The average of each area is used to plot a radar diagram that graphically presents the results.
One of the main limitations of this approach come from the fact that the author did not provide any reference material to aid the interpretation of the statements. 
Thus, the participants in the survey should assess the statement and assign a 5 point Likert scale to it only by analyzing the statement description. 
\\    

\noindent
\textbf{7. \textit{Agile Enterprise Survey}} \\
The approach named \textit{Agile Enterprise Survey} proposed in \cite{lewis2016-stormAgileEnterpriseSurvey} is a web-based self-assessment survey designed by Storm Consulting in collaboration with the Dresden University.
The survey's questionnaire has 44 statements and 2 open-ended questions organized in 6 sections as follows: 
\textit{Values and Practices}, 
\textit{Working Environment},  
\textit{Capabilities - human resources}, 
\textit{Activities},
\textit{``Blue sky'' thinking}, and
\textit{Organisation background}. 
The survey presents a set of statements and asks the participant to specify, using a 5-point Likert scale, how well these statements reflect their organization.
The section \textit{``Blue Sky'' thinking} presents two open-ended questions including one that asks ``\textit{If you could wave a magic wand to make any changes you wished to your working environment, what would you change?}''.
After submitting the answers, the assessment result is calculated and sent by email to the participant.
The authors did not indicate how the questions where selected, how they are linked with the agile practices and how the answers are analyzed.    
\\
 
\noindent
\textbf{8. \textit{Agile Excellerate}} \\
\textit{Agile Excellerate} \citep{sfirlogea2020-AgileExcellerate} is a web-based questionnaire that assists agile team members to evaluate their understanding of agile principles, their values and adherence to good practices.
It also highlights potential issues related to \textit{trust}, \textit{team cohesion}, \textit{commitment}, \textit{constructive conflicts} and \textit{accountability}. 
It is based on several theories related to 
self-organization 
(\textit{Container-Difference-Exchange} \citep{eoyang2001conditions}), 
team building 
(\textit{Five dysfunctions of a team} \citep{lencioni2012five}) 
and trust 
(\textit{Speed of trust} \citep{covey2006speed}).
The questionnaire is composed of 80 questions that assess the developer's opinions and perceptions about various aspects of the team agility. 
Questions are grouped in 8 analysis dimensions, covering the most important aspects of the Agile practice:
\textit{Respect and Communication},
\textit{Collaborative Improvement},
\textit{Sustainable Delivery},
\textit{Disciplined Self-Organization},
\textit{Predictable Quality},
\textit{Empowered Courage},
\textit{Focused Commitment}, and
\textit{Transparency and Visibility}.
Results are consolidated and analyzed by the authors (the current version of \textit{Agile Excellerate} does not feature automatic reporting). 
After analysis, the following reports are provided:
\begin{enumerate*}
    \item Radar chart of team agility based on all analysis dimensions;
    \item Results on each dimension emphasizing critical aspects (low scoring or abnormal distribution of answers)
    \item Correlation map (how various answers correlate or not)
    \item Container — Difference — Exchange score
    \item Scoring of potential team dysfunctions
    \item Trust analysis: integrity, intent, capabilities and results.
\end{enumerate*}
\\
   
\noindent
\textbf{9. \textit{Agile Health Dashboard}} \\
\textit{Agile Health Dashboard} \citep{lagestee2012-agile-health-dashboard} is a coaching tool available as a spreadsheet that helps agile teams to improve their development process continuously.
To use the tool, a team member should manually fill a pre-configured sheet entering information about each sprint (start and end dates, number of completed stories, team velocity, etc). 
Based on the data provided, the tool calculates and shows a dashboard organized in 4 areas: 
\textit{Sprint Planning}, 
\textit{Sprint Velocity}, 
\textit{Team Flow}, and 
\textit{Team Dynamics}.
The team should use the data emerging from the dashboard to find areas that need improvement.
\\

\noindent
\textbf{10. \textit{Agile Journey Index (AJI)}} \\
\cite{krebs2011-agile-journey-index} proposed an agility assessment model called \textit{Agile Journey Index (AJI)} that aids organizations in improving how they are applying agile practices. 
The model covers 19 key practices organized into 3 categories: 
\textit{Plan}, \textit{Do} and \textit{Feedback}. 
The assessment consists of rating each practice on a scale of 1 to 10. 
Although the model specifies criteria for each score, the evaluation of these criteria depends on qualitative analysis and there is no indication of how to identify the occurrence of these practices in real projects.
Another drawback of this model is that it considers only Scrum practices and neglects other agile methods.
\\

\noindent
\textbf{11. \textit{Agile Maturity Assessment}} \\
\cite{tousignant2019-AgileMaturityAssessment} presented the \textit{Agile Maturity Assessment}, a self-assessment approach available as a web tool that aims at measuring the organization agile maturity according to the \textit{Agile Maturity Model} \citep{tousignant2019-AgileMaturityMatrix}. 
The approach is composed of 60 agree/disagree statements that, after completed, generate a weighted total score that indicates the organization maturity level as follows:
0 - 80 points: ``\textit{Ad-hoc Agile}''; 
81-160 points: ``\textit{Doing Agile}'';
161-240 points: ``\textit{Being Agile}'';
241 - 320 points: ``\textit{Thinking Agile}''; and
$>$ 320 points: ``\textit{Culturally Agile}''.
One of the main limitations of the tool is the lack of transparency on how the total score is calculated.
The approach is a commercial tool but it is possible to run individual assessments free of charge.
\\

\noindent
\textbf{12. \textit{Agile Skills Self-Assessment}} \\
The BPM Institute proposed in \cite{BPMI2019-AgileSkillsSelf-Assessment} the \textit{Agile Skills Self-Assessment}, a web-based survey to assist agile team members in creating a professional development game plan. 
The survey covers 6 critical practice areas:
\textit{1. Agile Concepts},
\textit{2. Agile Rituals and Ceremonies},
\textit{3. Agile Business Analysis Principles},
\textit{4. Estimation and Velocity},
\textit{5. Creating and Managing Quality User Stories}, and
\textit{6. Utilizing Waterfall Business Analysis Techniques in Agile}.
Each area has 5 questions that should be scored using a 5-point Likert scale as follows:
\textit{1. Not at all},
\textit{2. Somewhat},
\textit{3. Middling},
\textit{4. Mostly}, and
\textit{5. Very}.
The final score, that ranges from 30 to 150, indicates the agile skill level
(\textit{Beginner},
\textit{Intermediate},
\textit{Advanced}, or 
\textit{Expert})
across the 6 critical practice areas covered.
\\

\noindent
\textbf{13. \textit{Agile Team Evaluation}} \\
In \cite{gunnerson2015-agileteamevaluation}, the author proposed a text-based questionnaire, \textit{Agile Team Evaluation}, to aid development teams to evaluate themselves. 
The questionnaire has 17 yes/no questions organized in 4 groups 
(\textit{Delivery of Business Value}, 
\textit{Code Health},
\textit{Team Health}, and 
\textit{Organization Health}).
The author, who intended to provide a ``\textit{less prescriptive approach}'', suggested questions such as ``\textit{Is the team healthy and happy?}'' and ``\textit{Is the code well architected?}'' that aim at promoting internal team discussions rather than defining a degree of agility to the team.
\\

\noindent
\textbf{14. \textit{Agility Maturity Self Assessment}} \\
\cite{maciver2010} proposed a self-assessment model named \textit{Agility Maturity Self-Assessment} that aims to identify the skills of individuals in six areas:
\textit{Agile Teams}, 
\textit{Agile Leadership}, 
\textit{Agile Project Management}, 
\textit{Agile Communication/Promotion}, 
\textit{Business Value}, and 
\textit{Risk Management}.
The questions have the following structure \textit{``How experienced are you in the given area...''}. The author did not provide any indication of how to analyze the answers.
\\

\noindent
\textbf{15. \textit{Agility Maturity Self Assessment Survey}} \\
In \cite{ribeiro2015}, the author proposed the \textit{Agile Maturity Self-Assessment Survey}, a survey where the participants can assess their skill in agile development by answering a questionnaire composed of 25 questions (including an open question). 
The author did not provide indications of how to analyze the answers and assess the skill of the individuals. 
\\

\noindent
\textbf{16. \textit{Agility Questionnaire}} \\
In \cite{britsch2017-agility-questionnaire}, \citeauthor{britsch2017-agility-questionnaire} proposed the \textit{Agility Questionnaire}, a spreadsheet-based questionnaire that helps to assess whether agile development is the proper approach for a specific organization and project and highlights associated challenges, risks and areas where specific tailoring is required.
The \textit{Agility Questionnaire} is not a self-assessment approach. 
Instead of that, the approach was designed to support consultant companies (called suppliers) and organizations willing to adopt agile development (called clients) to collaboratively assess the client's agile capability and propose the best ways to work.
The questionnaire is composed of 60 questions organized into two parts: 
\textit{Agility Profile} and \textit{Project Profile}.
Each question should be assigned with a 5-point agree/disagree Likert scale
(\textit{Strongly Agree}, \textit{Agree}, \textit{Neutral}, \textit{Disagree}, and \textit{Strongly Disagree}).
The questions of the first part (\textit{Agility Profile}) are organized into 6 areas:  
\textit{Value Focus},
\textit{Ceremony},
\textit{Collaboration},
\textit{Decisions and Information},
\textit{Responsiveness}, and
\textit{Experience}.
The questions of the second part (Project Profile) are organized into 12 areas:
\textit{Confidence},
\textit{Objectives and Goals},
\textit{Volatility},
\textit{Funding / Resourcing Challenge},
\textit{Analysis Challenge},
\textit{Political / Delivery Challenge},
\textit{Technology Challenge},
\textit{Design Challenge},
\textit{Reputational Risk},
\textit{Legal / Regulatory Risk},
\textit{Financial Risk}, and
\textit{Operational Risk}.
\\
\\

\noindent
\textbf{17. \textit{Back-of-a-Napkin Agile Assessment}} \\
The \textit{Back-of-a-Napkin Agile Assessment}  \citep{hendrickson2008-Back-of-a-NapkinAgileAssessment} is a text-based agile assessment checklist composed of 10 statements that aim at promoting internal team discussions rather than assessing the agility to the team.
\\

\noindent
\textbf{18. \textit{Balbes' Agility Assessment}} \\
\cite{balbes2015-HowAgileAreYou} proposed a text-based self-assessment approach to evaluate how agile teams are improving their ability to be agile over time.
The approach is composed of assessment questions that have 6 statements that correspond to different levels of maturity as described below:
\textit{Level 0: No Capability};
\textit{Level 1: Beginning};
\textit{Level 2: Learning};
\textit{Level 3: Practicing};
\textit{Level 4: Measuring}; and
\textit{Level 5: Innovating}
The assessment questions are grouped into 9 different areas:
\textit{Technical Craftsmanship},
\textit{Quality Advocacy},
\textit{User Experience},
\textit{Team Dynamics},
\textit{Product Ownership},
\textit{Project Management},
\textit{Risk Management},
\textit{Organizational Support}, and
\textit{Change Management}.
Once an assessment is complete and responses to each question are evaluated, results can be aggregated to show progress in each area.
\\

\noindent
\textbf{19. \textit{Borland Agile Assessment}} \\
In \cite{schumacher2009-borland}, \citeauthor{schumacher2009-borland} presented the \textit{Borland Agile Assessment 2009}, a text-based survey that was initially developed to be used as an internal coaching tool.
The survey consists of 12 questions answered on a 5-point agree/disagree scale.
There is no ``score'' to this assessment that should be administered anonymously with results reported in an aggregate form. 
The \textit{Borland Agile Assessment 2009} is a diagnostic tool to help development teams reflect on their processes and identify ways to improve (although the author did not make clear how to analyze the results).  
It should not be used to measure ``improvement'' from a previous assessment, only relative importance of potential improvements to their current situation.
The author claimed the approach was cross-referenced with the Manifesto for Agile Development \cite{beck2001}, as well as with agile principles from \cite{cockburn2002}, \cite{shore2007art}, and \cite{ambler2006}.
However, no evidence of such relations was provided.
\\

\noindent
\textbf{20. \textit{Business Agility Manifesto}} \\
\textit{Business Agility Manifesto} \citep{burlton2018-BusinessAgilityManifesto} is a text-based assessment approach that aims to provide initial insights into an organization’s need and readiness to become more agile. 
The approach consists of a questionnaire composed of 47 yes/no questions divided into 8 sections:
\textit{1. Perpetual Change},
\textit{2. Business Strategy and Value Creation},
\textit{3. Business Integrity},
\textit{4. Business Solution Agility},
\textit{5. Organization Agility},
\textit{6. Value Chain Perspective},
\textit{7. Business Knowledge and its Management}, and
\textit{8. Business Knowledge-Base / Single source of business truth}.
The authors suggested the survey can be used to help organizations understand their gaps to become more agile but failed to provide details on how to analyze the survey results. 
\\

\noindent
\textbf{21. \textit{Cargo Cult Agile Checklist}} \\
\textit{Cargo Cult Agile Checklist} \citep{wolpers2016-cargocultagilechecklist} is a text-based questionnaire that should be used as a start point for organizations adopting agile development to assess what part of the agile transition is going well and where action needs to be taken.
The questionnaire has 25 yes/no questions that are similar to the agile smells described in \cite{telemaco2020} in the sense they denote practices that may jeopardize the adoption of the agile development culture.
One of the main limitations of this approach is that the author only provided the question statement. There is no further description or hint on how to identify the occurrence of such practices.
Regarding the analysis of the results, the author provided a 5-level scale ranging from ``\textit{Well done!}'' (the first level with 0-2 ``yes'') to ``You either haven’t started going agile yet'' (the last level with 21-25 ``yes'').
\\
    
\noindent
\textbf{22. \textit{Comparative Agility (CA)}} \\
In \cite{williams2010ca}, \citeauthor{williams2010ca} proposed the \textit{Comparative Agility\texttrademark} (CA) method to aid organizations in determining their relative agile capability compared to other companies who responded to CA. 
The tool, which is available as a survey-based tool, assesses agility using 7 dimensions: 
\textit{Teamwork},
\textit{Requirements},
\textit{Planning}, 
\textit{Technical Practices}, 
\textit{Quality}, 
\textit{Culture}, and 
\textit{Knowledge Creation}.
Each dimension has between three and six characteristics (32 in total) and 
each characteristic comprises approximately four agile practices (125 in total).
For each practice, the respondent indicates the truth of the practice using a 5-point Likert scale:
\textit{True};
\textit{More true than false}; 
\textit{Neither true nor false}; 
\textit{More false than true}; or 
\textit{False}.
Although the approach uses an innovative assessment technique (by comparing the answers given by the company with a global trend), the authors neglected to indicate how the practices were identified, how they are related to the agile methods, and how they can be verified.
One of the questions that composes the method, for example, is 
\textit{``Team  members leave planning meetings knowing what needs to be done and have confidence they can meet their commitments''}. There are no clear criteria to check the occurrence of this practice.
\\

\noindent
\textbf{23. \textit{Comprehensive Agility Measurement Tool (CAMT)}} \\
\textit{Comprehensive Agility Measurement Tool (CAMT)} \citep{erande2008CAMT} is a text-based tool that supports the assessment of an organization's level of agility.
The approach proposes a unit measure, \textit{Comprehensive Agility Index (CAI)}, that indicates the level of agility on a scale of 1 to 5, where 1 means ``\textit{least agile}'' and 5 means ``\textit{highly agile}''.
To calculate this index, the approach uses a questionnaire that assesses 10 critical agility factors:
\begin{enumerate*}
    \item \textit{TAKT time};
    \item \textit{Plant Capacity};
    \item \textit{Inventory};
    \item \textit{Problem Solving};
    \item \textit{e-manufacturing};
    \item \textit{Continuous Improvement};
    \item \textit{Operational Flexibility};
    \item \textit{SMED / quick changeover};
    \item \textit{Internal Customer Satisfaction}; and
    \item \textit{Human Resource Management}.
\end{enumerate*}
Each critical factor should be assigned with a 5-point Likert scale that scores from 1 to 5 points.
\\

\noindent
\textbf{24. \textit{Depth of Kanban}} \\
\textit{Depth of Kanban} \citep{achouiantz2013-depth-of-kanban}, proposed by Achouiantz, is a graph-based coaching tool (not an evaluation or compliance tool) for assessing the depth of Kanban \citep{anderson2010kanban} adoption in an organization. 
The tool is available as an offline spider graph that is structured around the 7 Kanban principles: 
\textit{1. Visualize}, 
\textit{2. Limit Work in Progress}, 
\textit{3. Manage Flow}, 
\textit{4. Make Policies Explicit}, 
\textit{5. Implement Feedback Loops}, 
\textit{6. Improve}, and 
\textit{7. Effects}. 
Each axe has a different number of yes/no questions (the \textit{Limit Work in Progress} axe, for example, has 4 questions while the \textit{Visualize} axe has 13). 
The result of each axe (i.e., the level of agility) is denoted by the number of ``yes'' received.
Regarding the analysis of the results, the author divided the spider graph into four areas (represented by different colors):
\textit{Necessary for sustainable improvements} (red),
\textit{Improving Sustainably} (yellow),
\textit{Excellent} (light green), and 
\textit{Lean} (dark green).
The red area on the graph defines the minimal depth a team must reach in order to start improving on its own. While the team is ``in the red'' it cannot improve. 
The other colors indicate other ``levels'' of depth, the greener the better.
\\

\noindent
\textbf{25. \textit{Department of Defense Guide}} \\
The \textit{Defense Innovation Board Guide: Detecting Agile BS} \citep{dod2018-DIBGuide} is a text-based approach to provide guidance to program executives and acquisition professionals on how to detect software projects that are really using agile development versus those that are using waterfall or spiral development in agile clothing (\textit{``agile-scrum-fall''}).
The guide is divided into 4 sections:
(a) Section 1, named \textit{``Key flags that a project is not really agile''}, has 6 statements that may indicate whether a project is not using a process based on agile development;
(b) Section 2 describes a set of tools usually used by agile teams;
(c) Section 3 has a questionnaire organized into 5 subsections:
\textit{Questions to Ask Programming Teams} with 4 open questions,
\textit{Questions for Program Management} with 4 open questions,
\textit{Questions for Customers and Users} with 3 open questions, and
\textit{Questions for Program Leadership} with 6 open questions; and
(d) Section 4 has a graphical version of the questionnaire with a flow connected through yes/no questions that illustrates the path to a desired agile development process.
\\
    
\noindent
\textbf{26. \textit{Enterprise and Team Level Agility Maturity Matrix}} \\
The \textit{Enterprise and Team Level Agility Maturity Matrix} \citep{eliassengroup2013-enterprise-agility-maturity-matrix} is an agility assessment method available as a spreadsheet divided into two sections: one for describing the Organization and another for describing the Development Team. 
There are many agile indicators for each section (14 organizational indicators and 37 team indicators) and each indicator ranges from a `0' (impeded) to a `4' (ideal). 
For each cell in the matrix, there is a simple explanation of what it means to be at that level for that indicator. 
\\
 
\noindent
\textbf{27. \textit{Enterprise Business Agility Maturity Survey}} \\
In \cite{ribeiro2018-EnterpriseBusinessAgilityMaturitySurvey}, \citeauthor{ribeiro2018-EnterpriseBusinessAgilityMaturitySurvey} proposed the \textit{Enterprise Business Agility Maturity Survey}, a survey-based approach to support organizations measure their agility capability.
The survey is composed of 53 questions (including an open question) organized into 6 sections:
\textit{Leadership and Culture},
\textit{Lean Business and Portfolio Management},
\textit{Organisational Structure},
\textit{Agile Mindset and Methods},
\textit{Performance and Measurements}, and 
\textit{Make It Stick and Sustain}.
The author did not provide indications of how to analyze the answers and to assess the organization agility capability.  
\\  
  
\noindent
\textbf{28. \textit{Five Key Numbers to Assess Agile Engineering Practices}} \\
\cite{nielsen2011-FiveKeyNumbersToAssessYourAgileEngineeringPractices} 
proposed a text-based questionnaire to assess the team's agile engineering practices.
The questionnaire comprises 5 questions that should be scored using a gauge scale divided into three areas: green, yellow, and red. 
For example, the question 
``\textit{How many manual steps does it take to get a build into production?}'' 
has its gauge scale divided as follows:
0-1 steps: green;
2-9 steps: yellow; and 
9-15 steps: red.
The red area indicates the engineering practice has to be improved. 
The yellow area indicates the engineering practice is acceptable but could be improved. 
The green area indicates the engineering practice is well implemented.  
\\

\noindent
\textbf{29. \textit{GAO's Agile Assessment Guide}} \\
The U.S. Government Accountability Office (GAO) has published in \cite{gao2020AgileAssessmentGuide} the \textit{Agile Assessment Guide} to aid federal agencies, departments, and auditors in assessing an organization’s readiness to adopt Agile methods.
The guide contains 5 text-based checklists to assess specific areas of agile adoption:
\textit{1. Adoption of Agile Methods Checklist}: 24 statements organized into 9 areas;
\textit{2. Requirements Development Checklist}: 16 statements organized into 8 areas;
\textit{3. Contracting for an Agile Program Checklist}: 9 statements organized into 3 areas;
\textit{4. Agile and Program Monitoring and Control Checklist}: 9 statements organized into 3 areas; and
\textit{5. Agile Metrics Checklist}: 14 statements organized into 6 areas.
\\

\noindent
\textbf{30. \textit{How Agile are you? A 50 Point Test}} \\
\textit{How Agile are you? A 50 Point Test} \citep{finite2019-A50PointTest} is a web-based survey to help agile teams to determine how agile they are.
The survey is composed of 50 yes/no questions, allowing a team to arrive at a score out of 50 for each respondent.
Every team member of the agile team, including the Product Owner, testers, and managers have to honestly answer each statement. 
Once each team member has completed the 50 point test, add up the points for each respondent and average them to arrive at a total score for the team. 
Ideally, an agile team should have an average score greater than 40 points.
If a team's score is below 40, they should be looking to update their processes and team culture.
\\    
    
\noindent
\textbf{31. \textit{IBM DevOps Practices Self-Assessment }} \\
\textit{IBM DevOps Practices Self-Assessment} \citep{ibmDevOpsPracticesSelfAssessment} is an agility assessment approach available as a web application. The solution contains 15 questions divided into 4 areas:
\textit{Demographic},
\textit{Practices},
\textit{Strategies}, and 
\textit{Motivation}.
The authors claimed the tool can ``evaluate the state of an organization’s software delivery approach''. However, there are no indications of how the questions were formed, how the answers should be analyzed and how the results are related to agile practices.
\\
   
\noindent
\textbf{32. \textit{Joe's Unofficial Scrum Checklist}} \\
\cite{little2012-JoesUnofficialScrumChecklist} adapted the approach proposed by \cite{kniberg2012scrumchecklist} and proposed the \textit{Joe's Unofficial Scrum Checklist}, an approach to assist Scrum teams in assessing their agility. 
The approach, that should be used as basis for discussion preferably with the full team, has a checklist with 87 yes/no questions organized into 6 areas:
\textit{The Bottom Line},
\textit{Core Scrum},
\textit{Recommended},
\textit{Engineering	Practices},
\textit{Scaling}, and 
\textit{Positive Indicators}.
\\    

\noindent
\textbf{33. \textit{Karlskrona Test}} \\
\cite{seuffert2019-KarlskronaTestOnline} presented a self-assessment approach, named \textit{Karlskrona Test}, that was developed in 2008-2009 with companies in Sweden and Germany
to see how far an agile adoption came and to monitor progress over time. 
The test has 11 single-choice questions where each question has 4 options (2 of them score 0 and 2 score 1). 
The author suggested to submit the survey to all team members.
The final result is calculated according to the average amount of points for the whole team and ranges from \textit{Grade 1 - Waterfall} to \textit{Grade 5 - Agile}.
Although the author claimed this approach is an ``\textit{easy way to claim an organization is agile}'', there are no indications of how the questions relate to agile practices or empirical evidence to support this statement.
\\

\noindent
\textbf{34. \textit{Kanban Maturity Assessment}} \\
\textit{Kanban Maturity Assessment} \citep{chiva2019-KanbanMaturityAssessment}
is a web-based assessment approach that allows managers and Kanban coaches or consultants to help the teams evaluate and understand their progress and level of understanding of principles and practices of the Kanban Method \citep{anderson2010kanban}.
The \textit{Kanban Maturity Assessment} consists of 9 sections:
Section 1 is reserved for the assessment configuration;
Sections 2 to 6 focus on the 6 core Kanban practices, namely 
\textit{visualize}, 
\textit{limit WIP}, 
\textit{manage flow}, 
\textit{explicit policies}, 
\textit{feedback loops}, and \textit{improvement};
Section 8 assesses service and organizational effects of Kanban adoption;
Section 9 focuses on \textit{Fitness for Purpose}. To what extent the Service is servicing customer expectations.
Each section contains a set of statements that should be answered using a 5-point Likert scale: 
\textit{Strongly agree},
\textit{Agree},
\textit{Neutral},
\textit{Disagree}, and
\textit{Strongly disagree}.
\\
\\
  
\noindent
\textbf{35. \textit{Lean Agile Intelligence}} \\
The \textit{Lean Agile Intelligence} \citep{mccalla2016-LeanAgileIntelligence} is an assessment platform available as online questionnaires.
The approach provides the ability to customize out-of-the-box assessment templates or create new questionnaires from a question bank compiled from published works of agile specialists, framework reference guides, and collaborative feedback sessions with coaches.
The assessment results are presented as dashboards that aggregate team assessment results in a format that captures a holistic view of the organization's agility maturity and identifies patterns preventing the organizations from achieving their desired outcomes.
\\

\noindent
\textbf{36. \textit{Lean/Agile Depth Assessment Checklist A3}} \\
\cite{yeret2013LeanAgileDepthAssessmentChecklistA3} adapted the approach proposed by \cite{achouiantz2013-depth-of-kanban} and proposed the \textit{Lean/Agile Depth Assessment Checklist A3}, a graph-based coaching tool for evaluating the current agile capability of a team.
The tool is available as an offline spider graph that is structured around 7 perspectives: 
\textit{1. Visualize \& Manage the Flow};
\textit{2. Business Value Driven Development};
\textit{3. Individuals and Interactions Feedback Loops};
\textit{4. Engineering Practices};
\textit{5. Build and Deployment};
\textit{6. Empowered Teams and Individuals}; and
\textit{7. Improve}.
Each axe has a different number of yes/no questions (the \textit{Visualize \& Manage the Flow} axe, for example, has 15 questions while the \textit{Build and Deployment} axe has 6). 
The result of each axe (i.e., the level of agility) is denoted by the number of ``yes'' received.
Regarding the analysis of the results, the author divided the graph into 3 areas (represented by different colors):
(a) the red area indicates that the team has to improve its capability in this perspective;
(b) the yellow area indicates that the team has an acceptable capability on this perspective but some problems need to be addressed; and
(c) the green area indicates that the team has good capability in this perspective.
\\  
  
\noindent
\textbf{37. \textit{Lebow's Agile Assessment}} \\
\cite{lebow2018-versionOneAgileAssessment} presented an agility self-assessment approach composed of two elements: (a) a questionnaire and (b) a checklist;
The questionnaire contains 10 agility factors
(\textit{Team Communication},
\textit{User Accessibility},
\textit{Team Location},
\textit{Team Structure},
\textit{Delivery Frequency},
\textit{Measurement of Progress},
\textit{Ability to Change Direction},
\textit{Testing},
\textit{Planning Approach}, and
\textit{Process Philosophy})
that are rated from 1 to 5, where 1 being the least agile and 5 being the most agile.
The final score of the questionnaire, which is the sum of the rate assigned to each agility factor, should be analyzed as follows:
50 points: \textit{Agile maven};
40-49 points: \textit{Agilist all the way};
30-39 points: \textit{Agilist in training};
20-29 points: \textit{Closet agilist};
10-19 points: \textit{Thanks for taking the test}.
The checklist, that is named ``\textit{You might not be agile if...}'', has 10 statements that describe ``bad'' practices (i.e., practices that may impair the adoption of agile development).
One of the statements, for example, is 
``\textit{Your white boards are mostly white}''.  
\\

\noindent
\textbf{38. \textit{Measure.Team}} \\
\textit{Measure.team} \citep{albrecht2020-MeasureTeam} is a web-based self-assessment survey that aids agile teams tracking their progress and monitoring improvements over time.
The survey has 16 statements that should be scored using a 5-point Likert scale as follows:
\textit{Not at all/Not sure},
\textit{Rarely},
\textit{Sometimes},
\textit{Often}, and
\textit{Consistently}.
After submitting the questionnaire, the tool calculates and presents an overall score and, for each statement, its corresponding score and the following sections:
\textit{Why it is valuable to be consistent},
\textit{How to start},
\textit{How to improve},
\textit{How to sustain}, and 
\textit{Additional Resources}.
\\
  
\noindent
\textbf{39. \textit{Nokia Test}} \\
The \textit{Nokia Test} \citep{vodde2010-NokiaTest} for Scrum teams was developed originally by Bas Vodde at Nokia Siemens Networks in Finland and has been updated several times with the contribution of Jeff Sutherland.
The test is a self-assessment questionnaire organized in 10 agile areas: 
\textit{Iteration},
\textit{In-Sprint Testing},
\textit{Sprint Stories},
\textit{Product Owner},
\textit{Product Backlog},
\textit{Estimation},
\textit{Sprint Burndown},
\textit{Retrospective},
\textit{ScrumMaster}, and
\textit{Team}.
Each area has a set of statement that should be scored by each person of the evaluated team. 
The total score of an area ranges from 0 to 10 and, hence, the total score ranges from 0 to 100.
The team score is the average of the total score of each team member.
The authors does not provide any indication of how to analyze the team score.  
\\ 
  
\noindent
\textbf{40. \textit{Objectives Principles Strategies (OPS)}} \\
\cite{soundararajan2013-ops} proposed in his PhD Thesis the \textit{Objectives, Principles and Strategies Framework} (OPS), a framework that assists agility assessment by identifying 5 elements:
\begin{enumerate*}
    \item \textit{Objectives} of the agile development;
    \item \textit{Principles} that support the \textit{Objectives};
    \item \textit{Strategies} that implement the \textit{Principles};
    \item \textit{Linkages} that relate \textit{Objectives} to \textit{Principles}, and \textit{Principles} to \textit{Strategies}, and
    \item \textit{Indicators} for assessing the extent to which an organization supports the implementation and effectiveness of the \textit{Strategies}.
\end{enumerate*}  
\\    
    
\noindent
\textbf{41. \textit{Open Assessments}} \\
The \textit{Open Assessments} \citep{scrumOrg2020-OpenAssessments}
is a series of web-based questionnaires focuses on assessing someone's knowledge on specific areas of Scrum.
The series is composed of the following tests:
\textit{Scrum Open}: 30 questions to assess basic knowledge of Scrum;
\textit{Nexus Open}: 15 questions to assess basic understanding of the Nexus Framework;
\textit{Product Owner Open}: 15 questions to assess knowledge of the role of the Product Owner in Scrum;
\textit{Developer Open}: 30 questions to assess knowledge of development practices used across a Scrum Team;
\textit{Scrum with Kanban Open}: 15 questions to assess knowledge of practicing Professional Scrum with Kanban; and
\textit{Agile Leadership Open}: 10 questions to assess knowledge of Agile Leadership essentials.
\\    
    
\noindent
\textbf{42. \textit{Organizational Agility Self-Assessment}} \\
The \textit{Scaled Agile} initiative proposed 8 self-assessment approaches \citep{scaledagile-metrics}:
\begin{enumerate*}
    \item \textit{Business Agility Self-Assessment},
    \item \textit{Lean Portfolio Management Self-Assessment}, 
    \item \textit{Continuous Learning Culture Self-Assessment}, 
    \item \textit{Organizational Agility Self-Assessment}, 
    \item \textit{Enterprise Solution Delivery Self-Assessment}, 
    \item \textit{Lean Agile-Leadership Self-Assessment}, 
    \item \textit{Agile Product Delivery Self-Assessment}, and 
    \item \textit{Team and Technical Agility Self-Assessment}.
\end{enumerate*}
These approaches are available as spreadsheet questionnaires where the respondents should assign a 6-point Likert scale to each question:
\textit{True} (5 points),
\textit{More True than False} (4 points), 
\textit{Neither False nor True} (3 points),
\textit{More False than True} (2 points),
\textit{False} (1 point), and
\textit{Not Applicable} (0 point).
The approaches more related to this research are the 
\textit{Organizational Agility Self-Assessment} \citep{scaledagile-organizationalAgilitySelfAssessment} and 
the \textit{Team and Technical Agility Self-Assessment} \citep{scaledagile-teamAndTechnicalAgilitySelfAssessment}.
The \textit{Organizational Agility Self-Assessment} enables organizations to assess their proficiency in the \textit{Organizational Agility} \citep{scaledagile-organizationalAgility} competency that describes how Lean-thinking people and Agile teams optimize their business processes, evolve strategy with clear and decisive new commitments, and quickly adapt the organization as needed to capitalize on new opportunities.
This assessment approach is composed of 29 questions divided into 3 dimensions: 
\textit{Lean Thinking People and Agile Teams},
\textit{Lean Business Operations}, and
\textit{Strategy Agility}.
\\
\\

\noindent
\textbf{43. \textit{People 10 Team Assessment Approach}} \\
\cite{shoukath2012-AreYouReallyAgile} presented the \textit{People 10 Team Assessment Approach} a text-based assessment approach for organizations to benchmark the agile maturity of their teams.
The approach is composed of 24 engineering practices 
(for example \textit{Continuous integration}, \textit{Refactoring}, \textit{Build frequency}).
Each practice has 2 statements, one that best describes an \textit{`iterative'} team and another that best describes an \textit{`agile'} team.
Someone using the approach to assess a given team should mark, for each engineering practice, 1 point against the statement that best describes the team: \textit{`iterative'} or \textit{`agile'} team.
At the end, the assessed team has 2 scores, an \textit{`iterative'} team score and \textit{`agile'} team score. The greater score indicates the team's strongest capability.
\\

\noindent
\textbf{44. \textit{Perceptive Agile Measurement (PAM)}} \\
The method proposed by \cite{so2009-pam}, the \textit{Perceptive Agile Measurement} (PAM), is an agility assessment approach composed of 48 yes/no question organized into 8 agile areas, namely \textit{Iteration Planning},
\textit{Iterative Development},
\textit{Continuous Integration and Testing},
\textit{Stand-Up Meetings},
\textit{Customer Access},
\textit{Customer Acceptance Tests},
\textit{Retrospectives}, and
\textit{Collocation}.    
\\

\noindent
\textbf{45. \textit{Quick Self-Assessment of Your Organization's Agility}} \\
\cite{parry2009QuickSelf-Assessment} defined a questionnaire to aid organizations to self-assess their agility.
The questionnaire has 22 statements that should be scored using the following scale:
1 point if the statement is not true for the team;
3 points if the statement is somehow true for the team; and 
5 points if the statement is completely true for the team.
The author failed in describing how to analyze the final score.
\\

\noindent
\textbf{46. \textit{Retropoly}} \\
\textit{Retropoly} \citep{sfirlogea2017-Retropoly} is a game, based on the Monopoly game concept, to be used during retrospective meetings to aid agile teams self-assessing themselves.
It is mainly designed for Scrum teams, but it is suitable with minor adjustments for any other agile methodology. 
The game is an alternative to traditional retrospective meetings and some benefits reported by the authors are:
(a) It encourages an honest self-assessment of each member of the team and the positive feedback for the support provided by colleagues. It will improve the ability of the team to take common decisions in a timely manner and the practice of moderated debates;
(b) It strengthens the team relationships by getting to know each other through sharing of personal life aspects, like hobbies and passions; and
(c) It allows to observe how retrospectives are improving over the time.
The game contains, among other things, a deck of 18 cards with questions about agile practices.
\\

\noindent
\textbf{47. \textit{Scrum Assessment Series}} \\
The \textit{Scrum Assessment Series} \citep{hawks2013-ScrumAssessmentSeries} is an agility assessment approach divided into 5 series, each focusing on a different Scrum practice:
1. \textit{Daily Scrum};
2. \textit{Retrospective};
3. \textit{Sprint Planning};
4. \textit{Sprint Review}; and
5. \textit{Release Planning}.
Each series contains a questionnaire with yes/no questions organized into 3 sections (\textit{The Basics}, \textit{Good}, and \textit{Awesome}) and a section \textit{Ideas for Improvement} with statements to aid improving that area.
\\

\noindent
\textbf{48. \textit{Scrum Checklist}} \\
The \textit{Scrum Checklist} \citep{kniberg2012scrumchecklist} is a tool to help development teams getting started with Scrum, or assessing their current implementation of Scrum. 
The checklist is made up of 80 yes/no questions divided into 4 groups: 
\textit{The Bottom Line};
\textit{Core Scrum};
\textit{Recommended But Not Always Necessary};
\textit{Scaling}; and
\textit{Positive Indicators}.
According to the author, the items on the checklist are not rules and therefore were not designed to be verifiable or to produce a measure that indicates the level of compliance with Scrum. 
Instead, they are guidelines that might be used by teams as a discussion tool at the retrospective meetings. 
Examples of items on the checklist are 
\textit{``Whole team believes plan is achievable?''} 
or 
\textit{``Having fun? High energy level?''}.
\\

\noindent
\textbf{49. \textit{ScrumMaster Checklist}} \\
The \textit{ScrumMaster Checklist} \citep{james2007-scrummasterChecklist} is a text-based coaching tool for Scrum Masters elaborated according to the personal experience of the author that has a large experience training Scrum Masters.
The approach is divided into 2 parts. 
The first part contains 42 one-choice questions while the second part contains open questions to describe the organizational impediment. 
The questions in the first part are organized into 4 parts:
\begin{enumerate*}
    \item \textit{How Is My Product Owner Doing?}; 
    \item \textit{How Is My Team Doing?};
    \item \textit{How Are Our Engineering Practices Doing?}, and
    \item \textit{How Is The Organization Doing?}.
\end{enumerate*}
Each question should be marked with one of the following options:
\textit{Option 1} (if the respondent considering they are ``doing well'');
\textit{Option 2} (for ``could be improved and I know how to start'');
\textit{Option 3} (for ``could be improved, but how?''); or
\textit{N/A} (for ``not applicable'' or ``would provide no benefit'').
The author did not give indications on how to calculate and analyze the results.
The instructions provided in the approach indicate that if the respondents check off most of the items, 
they are on track to become an efficient Scrum Master.
\\   

\noindent
\textbf{50. \textit{Self Assessment Tool for Transitioning to Agile}} \\
\cite{rothman2013} proposed the \textit{Self Assessment Tool for Transitioning to Agile}, a self-assessment tool for measuring agile maturity composed of 8 questions.
The author also supplied, for some question, the expected answers for those organizations willing to adopt agile development and a discussion about the answers. 
As an example, the question: 
``\textit{If you are doing iterations, are they four weeks or less? The answer should be yes. Many of us like one or two week iterations. Why? Because you get feedback more often rather than less often. And, you get to see working software}''.   
\\

\noindent
\textbf{51. \textit{Squad Health Check Model}} \\
The \textit{Squad Health Check Model} \citep{kniberg2014-SquadHealthCheckModel} is a game-based approach used to aid organizations tracking the health of their squads (the term used by the authors to denote a small, cross-functional, and self-organizing development team). 
The model, that was firstly developed and applied at the company Spotify, prescribes three phases: 
\textit{Phase 1.} A Workshop to collect data where members of a squad discuss and assess their current situation based on a number of different perspectives, namely 
\textit{Delivery Value},
\textit{Easy to Release},
\textit{Fun},
\textit{Health of Codebase},
\textit{Learning},
\textit{Mission},
\textit{Pawns or Players},
\textit{Speed},
\textit{Suitable Process},
\textit{Support}, and 
\textit{Teamwork}.
This phase is supported by a deck of cards where each card has 2 statements, one green and one red. The green statement describes a good aspect of the assessed perspective and the red describes a bad aspect.
For each perspective, the team has to define a colour that best describes their current squad for that perspective where: 
(a) \textit{Green} means the squad is satisfied with their ability on that perspective and does not see need for improvement now;
(b) \textit{Yellow} means there are some important problems that need to be addressed; and
(c) \textit{Red} means the perspective needs to be improved.
\textit{Phase 2}. Create a graphical summary of the result.
\textit{Phase 3}. Use the data to help the squads improve.
\\

\noindent
\textbf{52. \textit{Team and Technical Agility Self-Assessment}} \\
\textit{Team and Technical Agility} \citep{scaledagile-teamAndTechnicalAgility} competency which is the collection of foundation practices on which Agile development is based (see approach 42).
This assessment approach has 34 questions divided into 3 dimensions:
\textit{Agile Teams},
\textit{Team of Agile teams}, and
\textit{Built-in Quality}.
Regarding the analysis of the results, the approaches calculate the average score for each dimension considering the questions with a positive score (\textit{Not applicable} answers are not counted in the final result) and present a radar chart with the dimensions and their corresponding score.
However, there is no clear indication of how to interpret these results and which action should be taken.
\\

\noindent
\textbf{53. \textit{Team Barometer}} \\
\textit{Team Barometer} \citep{janlen2014-Barometer} is an approach that has a twofold goal: (a) to evaluate how an agile team gets stronger over time, and (b) to be conducted as an alternative to the traditional iteration retrospective meetings.
The approach is executed as a survey in a workshop with the whole agile team. 
The survey consists of 16 team characteristics, packaged as a deck of cards.
Each card has a headline naming the corresponding characteristic of the team, a green and a red statement. 
The green statement denotes a good practice while the red statement denotes a bad practice.
For example, the card that corresponds to the characteristic ``Trust'' has the following statements: 
``\textit{We have the courage to be honest with each other. We don't hesitate to engage in constructive conflicts}'' (green) and 
``\textit{Members rarely speak their mind. We avoid conflicts. Discussions are tentative and polite.}'' (red).
Team members vote green, yellow or red for each card in the meeting. 
Green means that the member agrees with the green statement, red that the member agrees with the red statement. A yellow vote means that the member thinks it is neither green nor red but something in the middle.
Once all cards have been run through, the team reflects and discusses the results.
\\

\noindent
\textbf{54. \textit{TeamMetrics}} \\
The \textit{TeamMetrics} \citep{verwijs2017TeamMetrics} is a web-based survey proposed by Christiaan Verwijs that aims at helping agile teams improve by gathering data about key team factors such as \textit{team morale}, \textit{motivation}, \textit{happiness}, \textit{learning}, \textit{performance}, \textit{communication}, and \textit{leadership} and interpret the results with the help of benchmarks.
The survey has 10 statements that should be scored using a 19-point Likert scale where the lowest value means \textit{Very inaccurate} and the highest value means \textit{Very accurate}.
One of the statements that compose the survey, for example, is ``\textit{My job requires me to use a number of high level or complex skills}''.
\\
    
\noindent
\textbf{55. \textit{Test Maturity Card Game}} \\
The \textit{Test Maturity Card Game} \citep{schuurkes2017-TestMaturityCardGame} is game-based tool designed to help teams assess and improve their testing capability. 
The approach is supported by a card game that helps teams discuss and identify strengths and weaknesses in their process.
The model consists of a set of criteria organized in 6 different areas: 
\textit{1. Test Culture},
\textit{2. Context},
\textit{3. Trait},
\textit{4. Skills},
\textit{5. Processes}, and
\textit{6. Artefacts}.
The team uses a card game to identify the criteria most relevant to their context.
The approach is supported by a card game that is used to aid the teams identify the most relevant criteria to their context and to assess the teams ability in the selected criteria.    
\\    
    
\noindent
\textbf{56. \textit{The Agile Self-Assessment Game}} \\
\cite{linders2019-TheAgileSelf-AssessmentGame} presented the \textit{Agile Self-Assessment Game}, a game-based approach that assists teams to reflect on their own team interworking, discover how agile they are and decide what they can do to increase their agility. 
The game consists of a book and a deck of cards with statements on applying agile practices organized into 5 ``suits'':
52 \textit{Basic Agile} cards,
39 \textit{Scrum} cards,
52 \textit{Kanban} cards,
26 \textit{DevOps} cards, and
26 \textit{Business Agility} cards.
It is worth mentioning that the book is not only available in English, but there's also a Spanish edition. 
Additionally, the cards have been translated into several other languages, including Dutch, French, German, Italian, Polish, Czech, and many more, making the approach accessible to a wider audience.  
\\    
    
\noindent
\textbf{57. \textit{The Art of Agile Development}} \\
\cite{shore2007art} proposed a self-assessment survey that aims to help agile teams review and evaluate their approach to adopting agile development.
It focuses on 5 important aspects of agile development: 
\textit{Thinking},
\textit{Collaborating},
\textit{Releasing},
\textit{Planning}, and 
\textit{Developing}.
The approach is available as a text-based survey composed of 46 yes/no questions.
Each question has a specific weight that ranges from 3 to 75. 
The final score, that is calculated as the sum of each question, should be analyzed as follows:
75 points or less: ``\textit{immediate improvement required}'' (red);
75 to 96 points: ``\textit{improvement necessary}'' (yellow);
97, 98, or 99: ``\textit{improvement possible}'' (green); and 
100: ``\textit{no further improvement needed}''.
\\    

\noindent
\textbf{58. \textit{The Joel Test: 12 Steps to Better Code}} \\
\cite{spolsky2000-JoelTest} a self-assessment questionnaire for measuring the agile proficiency of a software team. 
The questionnaire has 12 yes/no questions. 
Each `yes' answer scores 1 point and the final score, which is the sum of the 12 questions, should be analyzed as follows: 
A score of 12 means the organization is perfect, 11 is tolerable, but 10 or lower means the organization has serious problems.
\\    

\noindent
\textbf{59. \textit{Visual Management Self-Assessment}} \\
The \textit{Visual Management Self-Assessment} \citep{hogan2017-VisualManagementSelf-Assessment} is a web-based self-assessment survey to aid organizations in identifying what techniques they are currently doing and finding the next steps for improvement.
The survey is useful as a baseline for measuring an organization's depth of Kanban adoption over time and also as a checklist of ideas for techniques to try.
The survey is organized into 3 areas: 
\textit{Clarity} of the work (position of the work, performance measures, and identification of problems), 
\textit{Controls} (over team capacity and commitments to stakeholders) and
\textit{Collaboration} (feedback mechanisms and team collaboration practices).
The tool sends to all participants a report on insights into the state of Kanban across the survey once there are enough responses.
\\   
    
\noindent
\textbf{60. \textit{Yodiz's Team Agility Self Assessment}} \\
The \textit{Yodiz's Team Agility Self Assessment} \citep{yodiz2017-TeamAgilitySelfAssessment} is a self-assessment survey available as a spreadsheet that can be used to support agile teams understanding of whether and to which extent they are applying agile practices.
The survey is composed of 37 questions organized into 8 agile areas 
(\textit{Team}, 
\textit{Backlog},
\textit{Daily Scrum},
\textit{Sprint},
\textit{Coding Practices},
\textit{Testing},
\textit{Business}, and 
\textit{Retrospective}).
Each question assesses whether the team is applying a specific agile practice and  is scored using the following scale:
\textit{0 points: Never},
\textit{1 point: Rarely},
\textit{2 points: Occasionally},
\textit{3 points: Often},
\textit{4 points: Very Often}, and
\textit{5 points: Always}.
The findings from the survey are illustrated in the form of a pie chart. 
The chart visualizes the overall progress and where the team needs to improve. 

\section{Conclusion}

In this technical report, we presented a list of the main approaches and tools focused on agility assessment. 
We analyzed 60 approaches and tools including text-based, graph-based, game-based, spreadsheet-based, and web-based approaches.

\vskip 0.2in

\bibliography{99-bib}

\begin{thebibliography}{79}
\providecommand{\natexlab}[1]{#1}
\providecommand{\url}[1]{\texttt{#1}}
\expandafter\ifx\csname urlstyle\endcsname\relax
  \providecommand{\doi}[1]{doi: #1}\else
  \providecommand{\doi}{doi: \begingroup \urlstyle{rm}\Url}\fi

\bibitem[Achouiantz(2013)]{achouiantz2013-depth-of-kanban}
Christophe Achouiantz.
\newblock \emph{{Depth of Kanban}}, 2013.
\newblock URL
  \url{http://leanagileprojects.blogspot.com/2013/03/depth-of-kanban-good-coaching-tool.html}.

\bibitem[Agile(2012{\natexlab{a}})]{scaledagile-metrics}
Scaled Agile.
\newblock \emph{Metrics - Scaled Agile Framework}, 2012{\natexlab{a}}.
\newblock URL \url{https://www.scaledagileframework.com/metrics/}.

\bibitem[Agile(2012{\natexlab{b}})]{scaledagile-organizationalAgility}
Scaled Agile.
\newblock \emph{Organizational Agility}, 2012{\natexlab{b}}.
\newblock URL
  \url{https://www.scaledagileframework.com/organizational-agility/}.

\bibitem[Agile(2012{\natexlab{c}})]{scaledagile-organizationalAgilitySelfAssessment}
Scaled Agile.
\newblock \emph{Organizational Agility Self-Assessment}, 2012{\natexlab{c}}.
\newblock URL \url{https://www.scaledagileframework.com/metrics/#PF9}.

\bibitem[Agile(2012{\natexlab{d}})]{scaledagile-teamAndTechnicalAgility}
Scaled Agile.
\newblock \emph{Team and Technical Agility}, 2012{\natexlab{d}}.
\newblock URL
  \url{https://www.scaledagileframework.com/team-and-technical-agility/}.

\bibitem[Agile(2012{\natexlab{e}})]{scaledagile-teamAndTechnicalAgilitySelfAssessment}
Scaled Agile.
\newblock \emph{Team and Technical Agility Self-Assessment},
  2012{\natexlab{e}}.
\newblock URL \url{https://www.scaledagileframework.com/metrics/#T4}.

\bibitem[Albrecht and Eddings(2020)]{albrecht2020-MeasureTeam}
Kelly Albrecht and Sean Eddings.
\newblock \emph{Measure.team}, 2020.
\newblock URL \url{https://measure.team/}.

\bibitem[Ambler and Lines(2012)]{ambler2012disciplined}
Scott~W Ambler and Mark Lines.
\newblock \emph{Disciplined agile delivery: a practitioner's guide to agile
  software delivery in the enterprise}.
\newblock IBM press, 2012.

\bibitem[Ambler(2006)]{ambler2006}
S.W. Ambler.
\newblock Survey says: agile works in practice.
\newblock \emph{Dr. Dobb's Journal}, 31\penalty0 (9):\penalty0 62--64, 2006.

\bibitem[Anderson(2010)]{anderson2010kanban}
David~J Anderson.
\newblock \emph{Kanban: successful evolutionary change for your technology
  business}.
\newblock Blue Hole Press, 2010.

\bibitem[Balbes(2015)]{balbes2015-HowAgileAreYou}
Mark Balbes.
\newblock \emph{{How Agile Are You? Let's Actually Measure It!}}, 2015.
\newblock URL
  \url{https://adtmag.com/articles/2015/12/15/balbes-agile-model-0-intro.aspx}.

\bibitem[Beck et~al.(2001)Beck, Beedle, van Bennekum, Cockburn, Cunningham,
  Fowler, Grenning, Highsmith, Hunt, Jeffries, Kern, Marick, Martin, Mellor,
  Schwaber, Sutherland, and Thomas]{beck2001}
Kent Beck, Mike Beedle, Arie van Bennekum, Alistair Cockburn, Ward Cunningham,
  Martin Fowler, James Grenning, Jim Highsmith, Andrew Hunt, Ron Jeffries, Jon
  Kern, Brian Marick, Robert~C. Martin, Steve Mellor, Ken Schwaber, Jeff
  Sutherland, and Dave Thomas.
\newblock {Manifesto for agile software development}.
\newblock http://www.agilemanifesto.org/, 2001.

\bibitem[Bonamassa(2018)]{bonamassa2018AgileAdoptionInterview}
Michael Bonamassa.
\newblock \emph{Agile Adoption Interview}, 2018.
\newblock URL \url{https://pladcloud.typeform.com/to/HjcVKG}.

\bibitem[BPMI(2019)]{BPMI2019-AgileSkillsSelf-Assessment}
BPM~Institute BPMI.
\newblock \emph{Agile Skills Self-Assessment}, 2019.
\newblock URL \url{https://www.bpminstitute.org/skills-self-assessments#}.

\bibitem[Britsch(2017)]{britsch2017-agility-questionnaire}
Marcel Britsch.
\newblock \emph{{Agility Questionnaire}}, 2017.
\newblock URL
  \url{https://thedigitalbusinessanalyst.co.uk/agility-questionnaire-130b03133b98}.

\bibitem[Burlton et~al.(2018)Burlton, Ross, and
  Zachman]{burlton2018-BusinessAgilityManifesto}
Roger~T. Burlton, Ronald~G. Ross, and John~A. Zachman.
\newblock \emph{Business Agility Manifesto}, 2018.
\newblock URL
  \url{https://busagilitymanifesto.org/accompaniments/supplements/diagnostics}.

\bibitem[Campbell and MacIver(2010)]{maciver2010}
Bryan Campbell and Robbie MacIver.
\newblock Agility maturity self assessment.
\newblock
  http://www.robbiemaciver.com/documents/presentations/A2010-Agile\%20Maturity\%20Self-Assessment.pdf,
  2010.
\newblock Accessed: 2019-06-30.

\bibitem[Chiva(2019)]{chiva2019-KanbanMaturityAssessment}
Gerard Chiva.
\newblock \emph{{Kanban Maturity Assessment}}, 2019.
\newblock URL \url{https://aktiasolutions.com/kanban-maturity-assessment/}.

\bibitem[Cockburn(2002)]{cockburn2002}
Alistair Cockburn.
\newblock \emph{{Agile software development}}.
\newblock Addison-Wesley Longman Publishing Co., Inc., Boston, MA, USA, 2002.
\newblock ISBN 0-201-69969-9.

\bibitem[Covey and Merrill(2006)]{covey2006speed}
Stephen~R Covey and Rebecca~R Merrill.
\newblock \emph{The speed of trust: The one thing that changes everything}.
\newblock Simon and schuster, 2006.

\bibitem[{Department of Defense DOD}(2018)]{dod2018-DIBGuide}
{Department of Defense DOD}.
\newblock \emph{DIB Guide: Detecting Agile BS}, 2018.
\newblock URL
  \url{https://media.defense.gov/2018/Oct/09/2002049591/-1/-1/0/DIB_DETECTING_AGILE_BS_2018.10.05.PDF}.

\bibitem[Eliassen-Group(2013)]{eliassengroup2013-enterprise-agility-maturity-matrix}
Eliassen-Group.
\newblock \emph{{Enterprise Agility Maturity Matrix}}, 2013.
\newblock URL
  \url{http://blog.eliassen.com/introducing-the-enterprise-agility-maturity-matrix}.

\bibitem[Eoyang(2001)]{eoyang2001conditions}
Glenda~Holladay Eoyang.
\newblock \emph{Conditions for self-organizing in human systems}.
\newblock Union Institute, 2001.

\bibitem[Erande and Verma(2008)]{erande2008CAMT}
Ameya~S Erande and Alok~K Verma.
\newblock Measuring agility of organizations-a comprehensive agility
  measurement tool (camt).
\newblock \emph{International journal of applied management and technology},
  6\penalty0 (3), 2008.

\bibitem[Finite(2019)]{finite2019-A50PointTest}
Finite.
\newblock \emph{{How Agile are you? A 50 Point Test}}, 2019.
\newblock URL \url{https://www.finite.com.au/blog/2019/08/how-agile-are-you/}.

\bibitem[Fraser et~al.(2006)Fraser, Boehm, J{\"a}rkvik, Lundh, and
  Vilkki]{fraser2006agile}
Steven Fraser, Barry Boehm, Jack J{\"a}rkvik, Erik Lundh, and Kati Vilkki.
\newblock How do agile/xp development methods affect companies?
\newblock In Pekka Abrahamsson, Michele Marchesi, and Giancarlo Succi, editors,
  \emph{Extreme Programming and Agile Processes in Software Engineering}, pages
  225--228, Berlin, Heidelberg, 2006. Springer Berlin Heidelberg.
\newblock ISBN 978-3-540-35095-8.

\bibitem[Gandomani et~al.(2013{\natexlab{a}})Gandomani, Zulzalil, Ghani, Azim,
  and Sultan]{gandomani2013important}
Taghi~Javdani Gandomani, Hazura Zulzalil, Abdul Ghani, Abdul Azim, and
  Abu~Bakar Sultan.
\newblock Important considerations for agile software development methods
  governance.
\newblock \emph{Journal of Theoretical \& Applied Information Technology},
  55\penalty0 (3):\penalty0 345--351, 2013{\natexlab{a}}.

\bibitem[Gandomani et~al.(2013{\natexlab{b}})Gandomani, Zulzalil, Ghani,
  Sultan, and Nafchi]{gandomani2013obstacles}
Taghi~Javdani Gandomani, Hazura Zulzalil, Abdul Azim~Abdul Ghani, Abu Bakar~Md
  Sultan, and Mina~Ziaei Nafchi.
\newblock Obstacles in moving to agile software development methods at a
  glance.
\newblock \emph{Journal of Computer Science}, 9\penalty0 (5):\penalty0 620,
  2013{\natexlab{b}}.

\bibitem[GAO(2020)]{gao2020AgileAssessmentGuide}
U.S. Government Accountability~Office GAO.
\newblock \emph{Agile Assessment Guide}, 2020.
\newblock URL \url{https://www.gao.gov/assets/710/709711.pdf}.

\bibitem[Gunnerson(2015)]{gunnerson2015-agileteamevaluation}
Eric Gunnerson.
\newblock \emph{{Agile Team Evaluation}}, 2015.
\newblock URL
  \url{https://docs.microsoft.com/en-us/archive/blogs/ericgu/agile-team-evaluation}.

\bibitem[Hawks(2013)]{hawks2013-ScrumAssessmentSeries}
David Hawks.
\newblock \emph{Scrum Assessment Series}, 2013.
\newblock URL \url{https://agilevelocity.com/scrum-assessment-series/}.

\bibitem[Hendrickson(2008)]{hendrickson2008-Back-of-a-NapkinAgileAssessment}
Elisabeth Hendrickson.
\newblock \emph{Back-of-a-Napkin Agile Assessment}, 2008.
\newblock URL
  \url{https://testobsessed.com/2008/11/back-of-a-napkin-agile-assessment/}.

\bibitem[Hermida(2009)]{hermida2009abetterteam}
Sebastian Hermida.
\newblock {A better team}.
\newblock http://www.abetterteam.org/, 2009.
\newblock Accessed: 2019-06-30.

\bibitem[Hoffmann et~al.(2018)Hoffmann, Bona, and
  Petit]{hoffmann2018-AgileAlert}
Rainer Hoffmann, Tilman Bona, and Serge Petit.
\newblock \emph{Agile Alert}, 2018.
\newblock URL \url{https://huz.de/en/agile-alert/}.

\bibitem[Hogan(2017)]{hogan2017-VisualManagementSelf-Assessment}
Ben Hogan.
\newblock \emph{Visual Management Self-Assessment}, 2017.
\newblock URL \url{http://agileben.com/blog/kanban-online-assessment}.

\bibitem[IBM(2008)]{ibmDevOpsPracticesSelfAssessment}
IBM.
\newblock {IBM devops self-assessment}.
\newblock https://devopsassessment.mybluemix.net/, 2008.
\newblock Accessed: 2019-06-30.

\bibitem[James(2007)]{james2007-scrummasterChecklist}
Michael James.
\newblock \emph{ScrumMaster Checklist}, 2007.
\newblock URL
  \url{https://scrummasterchecklist.org/pdf/ScrumMaster_Checklist_12_unbranded.pdf}.

\bibitem[Janlén(2014)]{janlen2014-Barometer}
Jimmy Janlén.
\newblock \emph{Team Barometer}, 2014.
\newblock URL
  \url{https://blog.crisp.se/2014/01/30/jimmyjanlen/team-barometer-self-evaluation-tool}.

\bibitem[Kniberg(2012)]{kniberg2012scrumchecklist}
Henrik Kniberg.
\newblock Scrum checklist.
\newblock http://www.crisp.se/scrum/checklist, 2012.
\newblock Accessed: 2019-06-30.

\bibitem[Kniberg and Lindwall(2014)]{kniberg2014-SquadHealthCheckModel}
Henrik Kniberg and Kristian Lindwall.
\newblock \emph{Squad Health Check Model}, 2014.
\newblock URL
  \url{https://engineering.atspotify.com/2014/09/16/squad-health-check-model/}.

\bibitem[Krebs(2011)]{krebs2011-agile-journey-index}
Bill Krebs.
\newblock \emph{Agile Journey Index}, 2011.
\newblock URL \url{http://www.agiledimensions.com/blog/}.

\bibitem[Lagestee(2012)]{lagestee2012-agile-health-dashboard}
Len Lagestee.
\newblock \emph{Agile Health Dashboard}, 2012.
\newblock URL
  \url{http://illustratedagile.com/2012/09/25/how-to-measure-team-agility/}.

\bibitem[Lebow(2018)]{lebow2018-versionOneAgileAssessment}
Jodi Lebow.
\newblock \emph{{Agile Assessment: Test Your Team's Agility}}, 2018.
\newblock URL \url{https://digital.ai/resources/agile-101/agile-assessment}.

\bibitem[Lencioni(2012)]{lencioni2012five}
Patrick Lencioni.
\newblock The five dysfunctions of a team.
\newblock In \emph{A workshop for teams. Pfeiffer, a Wiley Imprint}, 2012.

\bibitem[Lewis and Wendler(2016)]{lewis2016-stormAgileEnterpriseSurvey}
Russ Lewis and Roy Wendler.
\newblock \emph{Agile Enterprise Survey}, 2016.
\newblock URL \url{http://www.storm-consulting.com/agile-enterprise-survey/}.

\bibitem[Linders(2019)]{linders2019-TheAgileSelf-AssessmentGame}
Ben Linders.
\newblock \emph{The Agile Self-assessment Game}.
\newblock 2019.
\newblock URL \url{https://www.benlinders.com/game/}.

\bibitem[Little(2012)]{little2012-JoesUnofficialScrumChecklist}
Joseph~H. Little.
\newblock \emph{{Joe's Unofficial Scrum Checklist}}, 2012.
\newblock URL
  \url{http://agileconsortium.pbworks.com/w/file/fetch/66642311/Joe\%E2\%80\%99s\%20Unofficial\%20Scrum\%20CheckList\%20V13.pdf}.

\bibitem[McCalla and Gifford(2016)]{mccalla2016-LeanAgileIntelligence}
Michael McCalla and James Gifford.
\newblock \emph{Lean Agile Intelligence}, 2016.
\newblock URL \url{https://www.leanagileintelligence.com/}.

\bibitem[Nielsen(2011)]{nielsen2011-FiveKeyNumbersToAssessYourAgileEngineeringPractices}
Jeff Nielsen.
\newblock \emph{Five key numbers to gauge your agile engineering efforts},
  2011.
\newblock URL
  \url{https://www.slideshare.net/jeffreymads/five-key-numbers-to-gauge-your-agile-engineering-efforts-7980346}.

\bibitem[Nowinski(2016)]{nowinski2016-agileassessment}
Piotr Nowinski.
\newblock \emph{Agile Assessment}, 2016.
\newblock URL \url{http://piotr-nowinski.pl/agile-assessment/}.

\bibitem[Parry(2009)]{parry2009QuickSelf-Assessment}
Charles Parry.
\newblock \emph{Quick self-assessment of your organization’s agility}, 2009.
\newblock URL
  \url{http://www.signetconsulting.com/action_items/assessment.php}.

\bibitem[Qumer and Henderson-Sellers(2008)]{qumer2008-agile-adoption}
Asif Qumer and Brian Henderson-Sellers.
\newblock A framework to support the evaluation, adoption and improvement of
  agile methods in practice.
\newblock \emph{Journal of Systems and Software}, 81\penalty0 (11):\penalty0
  1899--1919, 2008.

\bibitem[Ribeiro(2015)]{ribeiro2015}
Eduardo Ribeiro.
\newblock {Agility maturity self assessment survey}.
\newblock
  https://beyondleanagile.com/2015/12/08/agile-maturity-self-assessment-survey-published-at-scrumalliance/,
  2015.
\newblock Accessed: 2019-06-30.

\bibitem[Ribeiro(2018)]{ribeiro2018-EnterpriseBusinessAgilityMaturitySurvey}
Eduardo Ribeiro.
\newblock \emph{Enterprise Business Agility Maturity Survey}, 2018.
\newblock URL
  \url{https://beyondleanagile.com/2018/12/12/enterprise-business-agility-maturity-survey/}.

\bibitem[Rothman(2013)]{rothman2013}
Johanna Rothman.
\newblock \emph{Self assessment tool for transitioning to agile}, 2013.
\newblock URL
  \url{https://www.jrothman.com/mpd/agile/2013/04/self-assessment-tool-for-transitioning-to-agile/}.
\newblock Accessed: 2019-06-30.

\bibitem[Sahota(2012)]{sahota2012-agile-adoption}
Michael Sahota.
\newblock \emph{An Agile Adoption and Transformation Survival Guide}.
\newblock Lulu. com, 2012.

\bibitem[Schoots and Schuurkes(2017)]{schuurkes2017-TestMaturityCardGame}
Huib Schoots and Joep Schuurkes.
\newblock \emph{Test Maturity Card Game}, 2017.
\newblock URL
  \url{https://www.huibschoots.nl/wordpress/wp-content/uploads/2017/02/Test-Improvement-Huib-Schoots-Joep-Schuurkes.pdf}.

\bibitem[Schumacher(2009)]{schumacher2009-borland}
Dale Schumacher.
\newblock \emph{Borland Agile Assessment 2009}, 2009.
\newblock URL
  \url{https://borland.typepad.com/agile_transformation/2009/03/borland-agile-assessment-2009.html}.

\bibitem[Scrum.Org(2020)]{scrumOrg2020-OpenAssessments}
Scrum.Org.
\newblock \emph{Open Assessments}, 2020.
\newblock URL \url{https://www.scrum.org/open-assessments}.

\bibitem[Seuffert(2019)]{seuffert2019-KarlskronaTestOnline}
Mark Seuffert.
\newblock \emph{Karlskrona test online}, 2019.
\newblock URL \url{https://mayberg.se/karlskrona-test-online}.

\bibitem[Sfirlogea and Georgescu(2017)]{sfirlogea2017-Retropoly}
Sorin Sfirlogea and Florian Georgescu.
\newblock \emph{Retropoly}, 2017.
\newblock URL \url{https://www.agilepractice.eu/retropoly/}.

\bibitem[Sfirlogea and Georgescu(2020)]{sfirlogea2020-AgileExcellerate}
Sorin Sfirlogea and Florian Georgescu.
\newblock \emph{Agile Excellerate}, 2020.
\newblock URL \url{https://www.agilepractice.eu/agile-excellerate/}.

\bibitem[Shore and Warden(2007)]{shore2007art}
James Shore and Shane Warden.
\newblock \emph{{The art of agile development: pragmatic guide to agile
  software development}}.
\newblock " O'Reilly Media, Inc.", 2007.

\bibitem[Shoukath(2012)]{shoukath2012-AreYouReallyAgile}
Nisha Shoukath.
\newblock \emph{Are you really agile?}, 2012.
\newblock URL
  \url{https://blog.people10.com/are-you-really-agile-a-free-assessment-here/}.

\bibitem[So and Scholl(2009)]{so2009-pam}
Chaehan So and Wolfgang Scholl.
\newblock Perceptive agile measurement: New instruments for quantitative
  studies in the pursuit of the social-psychological effect of agile practices.
\newblock In Pekka Abrahamsson, Michele Marchesi, and Frank Maurer, editors,
  \emph{Agile Processes in Software Engineering and Extreme Programming}, pages
  83--93, Berlin, Heidelberg, 2009. Springer Berlin Heidelberg.
\newblock ISBN 978-3-642-01853-4.

\bibitem[Soundararajan(2013)]{soundararajan2013-ops}
Shvetha Soundararajan.
\newblock \emph{Assessing agile methods: investigating adequacy, capability,
  and effectiveness (an objectives, principles, strategies approach)}.
\newblock PhD thesis, Virginia Tech, 2013.

\bibitem[Spolsky(2000)]{spolsky2000-JoelTest}
Joel Spolsky.
\newblock \emph{The Joel Test: 12 Steps to Better Code}, 2000.
\newblock URL
  \url{https://www.joelonsoftware.com/2000/08/09/the-joel-test-12-steps-to-better-code/}.

\bibitem[Telemaco et~al.(2020)Telemaco, Oliveira, Alencar, and
  Cowan]{telemaco2020}
Ulisses Telemaco, Toacy Oliveira, Paulo Alencar, and Don Cowan.
\newblock {A Catalogue of Agile Smells for Agility Assessment}.
\newblock \emph{IEEE Access}, 8:\penalty0 79239--79259, 2020.

\bibitem[Tousignant(2019{\natexlab{a}})]{tousignant2019-AgileMaturityAssessment}
Daniel Tousignant.
\newblock \emph{Agile Maturity Assessment}, 2019{\natexlab{a}}.
\newblock URL \url{https://capeprojectmanagement.com/individual-assessment/}.

\bibitem[Tousignant(2019{\natexlab{b}})]{tousignant2019-AgileMaturityMatrix}
Daniel Tousignant.
\newblock \emph{Agile Maturity Matrix}, 2019{\natexlab{b}}.
\newblock URL \url{https://capeprojectmanagement.com/agile-self-assessment/}.

\bibitem[VersionOne(2020)]{VersionOne2020}
VersionOne.
\newblock {The 14th annual state of agile report\textsuperscript{TM} 2020}.
\newblock https://stateofagile.com/, 2020.
\newblock Accessed: 2020-09-25.

\bibitem[Verwijs(2017)]{verwijs2017TeamMetrics}
Christiaan Verwijs.
\newblock \emph{TeamMetrics}, 2017.
\newblock URL \url{https://teammetrics.theliberators.com/}.

\bibitem[Vodde and Sutherland(2010)]{vodde2010-NokiaTest}
Bas Vodde and Jeff Sutherland.
\newblock \emph{Nokia Test}, 2010.
\newblock URL
  \url{https://www.scruminc.com/official-scrumbutt-test-otherwise-known/}.

\bibitem[Waters(2008)]{waters2008}
Kelly Waters.
\newblock \emph{{How agile are you? ((Take This 42 Point Test))}}, 2008.
\newblock URL
  \url{https://www.101ways.com/2008/01/21/how-agile-are-you-take-this-42-point-test/}.
\newblock Accessed: 2020-11-01.

\bibitem[Williams et~al.(2010)Williams, Rubin, and Cohn]{williams2010ca}
Laurie Williams, Kenny Rubin, and Mike Cohn.
\newblock Driving process improvement via comparative agility assessment.
\newblock In \emph{2010 Agile Conference}, pages 3--10. IEEE, 2010.

\bibitem[Wolpers(2016)]{wolpers2016-cargocultagilechecklist}
Stefan Wolpers.
\newblock \emph{Cargo Cult Agile Checklist}, 2016.
\newblock URL
  \url{https://age-of-product.com/cargo-cult-agile-state-agile-checklist-organization/}.

\bibitem[Yatzeck(2012)]{yatzeck2012}
Elena Yatzeck.
\newblock A corporate agile 10-point checklist.
\newblock
  http://pagilista.blogspot.com/2012/12/a-corporate-agile-10-point-checklist.html,
  Dec 2012.
\newblock Accessed: 2019-06-30.

\bibitem[Yeret(2013)]{yeret2013LeanAgileDepthAssessmentChecklistA3}
Yuval Yeret.
\newblock \emph{Lean/Agile Depth Assessment Checklist A3}, 2013.
\newblock URL
  \url{https://www.slideshare.net/yyeret/leanagile-depth-assessment}.

\bibitem[Yodiz(2017)]{yodiz2017-TeamAgilitySelfAssessment}
Yodiz.
\newblock \emph{Team Agility Self Assessment}, 2017.
\newblock URL
  \url{https://www.yodiz.com/blog/how-agile-is-your-team-take-our-team-agility-self-assessment-to-find-out/}.

\end{thebibliography}

\end{document}